\documentclass[aps,pra,twocolumn,groupedaddress,showkeys,showpacs]{revtex4-1}
\usepackage{graphicx}
\usepackage{dcolumn}
\usepackage{graphics}
\usepackage{epstopdf}
\usepackage{amssymb}
\usepackage{amsmath}
\usepackage{bm}
\usepackage{float}

\begin{document}
\title{Experimental observation of symmetry protected bound state in the continuum in a chain of dielectric disks}
\author{M.A. Belyakov$^1$, M.A. Balezin$^1$, Z.F. Sadrieva$^1$, P.V. Kapitanova$^1$, E.A. Nenasheva$^2$, A.F. Sadreev$^3$, A.A. Bogdanov$^{1}$}
\affiliation{$^1$Department of Photonics and Metamateials ITMO  University  St. Petersburg  197101,  Russia\\
$^2$Giricond Research Institute, Ceramics Co., Ltd., St. Petersburg 194223, Russia\\
$^3$Kirensky Institute of Physics Federal Research Center KSC SB RAS 660036 Krasnoyarsk Russia}
\date{\today}
\begin{abstract}
Existence of bound states in the continuum (BIC) manifests a general wave phenomenon firstly predicted in quantum mechanics in 1929 by J. von Neumann and E. Wigner. Today it is being actively explored in photonics, radiophysics, acoustics, and hydrodynamics. In this paper, we report the first experimental observation of electromagnetic bound state in the radiation continuum in 1D array of dielectric particles. By measurement of the transmission spectra of the ceramic disk chain at GHz frequencies we demonstrate how a resonant state in the vicinity of the center of the Brillouin zone turns into a symmetry-protected BIC with increase the number of the disks. We estimate a number of the disks when the radiation losses become negligible in comparison to material absorption and, therefore, the chain could be considered practically as infinite. The presented analysis is supplemented by measurements of the near fields of the symmetry protected BIC. All measurements are in a good agreement with the results of numerical simulation and analytical model based on tight-binding approximation. The obtained results provide useful guidelines for practical implementations of structures with bound states in the continuum that opens up new horizons for the development of optical and radiofrequency metadevices.

%quadratic growth and following saturation of Q-factor corresponding to the dominant contribution of the material losses

%We report experimental observation of symmetry protected bound states in the radiation continuum (BICs)  in the periodical array of ceramic disks predicted in $\Gamma$ point [Phys. Rev. A 96, 013841 (2017)]. Because of finite number of disks the BIC slightly leakages into radiation continuum that allows to observe the BIC in the transmission spectra by application of antennas at the ends of stack. Measured profiles of electromagnetic field and frequencies of the symmetry protected BIC at the $\Gamma$-point well agree  with results of COMSOL simulations. Moreover we exploit simple tight-binding model coupled with wide waveguide and wires that well describes the measured transmission spectra.

\end{abstract}
 \maketitle
%%%%%%%%%%%%%%%%%%%%%%%%%%%%%%%%%%%%%%%%%%%%%%%%%%%%%%%%%%%%%%%%%%%%%%%%%%%%%%%%%%%%%%%%%%%%%%%%%%%%%%
\section{Introduction}
%%%%%%%%%%%%%%%%%%%%%%%%%%%%%%%%%%%%%%%%%%%%%%%%%%%%%%%%%%%%%%%%%%%%%%%%%%%%%%%%%%%%%%%%%%%%%%%%%%%%%%%%

It is well known that a dielectric rod or slab supports waveguide modes formed under the condition of total internal reflection from the waveguide boundaries~\cite{adams1981introduction}. The wave numbers of the waveguide modes lie under the light line of the surrounding space making them orthogonal to the radiation continuum. When the dispersion curve crosses the frequency cut-off, the waveguide mode turns into a leaky mode  (resonant state)~\cite{adams1981introduction}. However, recently it was acknowledged that introduction of periodic modulation of the refractive index along the axis of the rod or slab discretises the radiation continuum, and this could result in complete suppression of radiation losses for leaky modes~\cite{torres2011twisted, Tikhodeev2005, Yablonskii2001, PRA2014,PRA92,Bulgakov2016}.
Therefore, the resonant state becomes localized, i.e. totally decoupled from the radiation continuum.
%The nature of this localizations is simple. In periodic systems, the leakage could occur only through the open diffraction channels. The number of open diffraction channels is defined   The amplitude of a diffraction channel is defined as the Fourier coefficient of the periodic part of the mode given      Intensity of radiation losses into a diffraction channel is proportional to the Fourier coefficient  However, under specific condition, the Fouri
Such localized solutions are known as {\em bound states in the continuum} (BICs) \cite{Neumann,Fonda}.
%It is well known that a dielectric rod or slab supports waveguide modes provided their spectrum lies below the light line with nonzero propagation wave number. That has transparent physical reason related to a phenomenon of full inner reflection \cite{Jackson}. At once the spectrum exceeds the light line the waveguide mode become leaky. However recently it was acknowledged that introduction of periodic modulation of the refractive index along the rod's axis brings important aspect into radiation continuum, a discretization that enormously suppresses a leakage.    With a certain set of parameters the leakage completely is blocked and the resonant state is localized decoupled from the outgoing wave in the ambient medium. Such localized solutions are known as bound states in the continuum (BICs) \cite{Neumann,Fonda,Stillinger}.
Recently, the immense progress in handling photonic crystals encouraged extensive studies on BICs
in various periodic photonic structures
\cite{Shipman,Marinica,HsuNature,Weimann,Wei,Yang,PRA2014,HuLu,Song,Zou,ZWang,YuanLu,BoZhen0,LiYin,Sadrieva2016,Ni,YWang,Gao,Blanchard}.
These studies are predominantly motivated by potential applications to resonant enhancement \cite{Magnusson, Zhang, Mocella, Magnusson2015}, lasing \cite{Kante,Bahari},  filtering of light \cite{Foley,Cui} and biosensing \cite{Romano,Liu}.

Among variety of the considered designs of photonic structures, the one-dimensional arrays of spheres or disks are distinctive because of the rotational symmetry. It gives rise to existence of BICs with orbital angular momentum. In the scattered spectra such a state manifest itself as a scattered field with orbital angular momentum travelling along the array \cite{PRA2014,OL,YuanLu,HuLuJOSAB}. This could be used for generation of {\em twisted light} and, therefore, for optomechanical manipulations~\cite{padgett2011tweezers}, quantum cryptography, and other applications~\cite{torres2011twisted}. Theory of BICs in the one-dimensional arrays of spheres and disks is developed  in Refs.~\cite{PRA92,PRA96} but, in spite of a variety of potential applications, experimental study is not presented for today.

%Among these different systems the one-dimensional arrays of spheres and disks are unique because of rotational symmetry  \cite{PRA92,PRA96}.

%(ii) they are easier to fabricate

In this work we report first experimental study of BICs in 1D axially symmetric array of dielectric particles.
We study transformation of the resonant state into symmetry protected BIC with increase of the number of the scatterers by measurement of
the transmission characteristic of the chain. In order to make the analysis more reliable, we measure the field profiles, which
confirm observation of the symmetry protected BIC.

\section{The resonant states in periodic array of dielectric disks}

\subsection{General theory of BIC in 1D chain}

The basic theory of BICs in the periodic stack of infinite number of dielectric disks is presented in Ref.~\cite{PRA96}.
Here, we briefly give the fundamentals of BIC theory and introduce the basic notations used further. According to Bloch theorem,
electric field $\mathbf{E}$ of eigenmodes in 1D periodic chain [see Fig.~\ref{fig1}(a)]  has the following form:
\begin{eqnarray}
      \mathbf{E}=\mathbf{U}_{m,n,k_z}(z,r)  e^{-i\omega t \pm ik_zz\pm im\varphi}.
\end{eqnarray}
Here $m$ is the orbital angular momentum (OAM), $k_z$ is the Bloch wavevector defined in the first Brillouin zone, and $n$ encodes the
index of the photonic band. The sign $\pm$ reflects the degeneracy of the modes propagating in opposite directions along the array and rotating in clockwise/counter-clockwise directions. Each mode with indices $m$ and $n$ has own dispersion $\omega=\omega_{m,n}(k_z)$. The dispersion of two lowest modes with $m=0$ is shown in Fig.~\ref{fig1}(b) with the red solid lines. The function  $\mathbf{U}_{m,n,k_z}(r,z)$ is a periodic function of the variable $z$ with a period $L$. Its expansion into the Fourier series
can be written as follows
\begin{eqnarray}
      \mathbf{U}_{m,n,k_z}(z,r)=\sum_n  \mathbf{C}_{m,n,k_z}(r) e^{\frac{2\pi i n}{L}z}.
\end{eqnarray}
In the general case, each mode consists of near fields and outgoing cylindrical waves. Therefore, the Fourier coefficient has the following
asymptotic far from the chain axis
\begin{eqnarray}
       \mathbf{C}_{m,n,k_z}(r) \rightarrow  \mathbf{C}_{m,n,k_z}^\infty H_m^{(1)}(\alpha_nr).
\end{eqnarray}
Here $H_m^{(1)}$ are the Hankel functions and
$%\begin{equation}
\alpha_n=\left[(\omega/c)^2-(k_z+2\pi n/L)^2\right]^{1/2}
$%\end{equation}
is the radial component of the wave vector. Indices $n$ numerate the diffraction channels. For open channels, $\alpha_n$ is real,
and for the closed ones, $\alpha_n$ is imagine.  The number of open diffraction channels, $\Lambda$, depends only on $\omega$ and $k_z$
[see Fig.~\ref{fig1}(b)]. If the amplitudes of all open diffraction channels are zero we have a BIC. In the case of
$k_z<\omega/c<|k_z\pm 2\pi /L|$, only one diffraction channel is open  [the yellow area in Fig.~\ref{fig1}(b)] and the radiation losses are determined only by the zero-order Fourier coefficient $\mathbf{C}_{m,0,k_z}(r)=\langle \mathbf{U}_{m,0,k_z}(r,z)\rangle_z$. Here $\langle\cdot\rangle_z$ implies integration over the period of the chain. The structures symmetric with respect to transformation $z\rightarrow-z$ allow for existence antisymmetric solutions at the $\Gamma$-point,
so-called, {\em symmetry protected} BIC. Protection by the symmetry means, for example, that a substrate with  not large refractive index (i.e. not opening the additional diffraction channels) does not destroy the BIC in spite of breaking the rotational symmetry of the chain. In other cases, BICs are called {\em accidental} or
off-$\Gamma$ BIC.

\subsection{Infinite chain \label{sec:inf_chain}}

\begin{figure}[t]%[htbp]
\includegraphics[width=0.9\columnwidth]{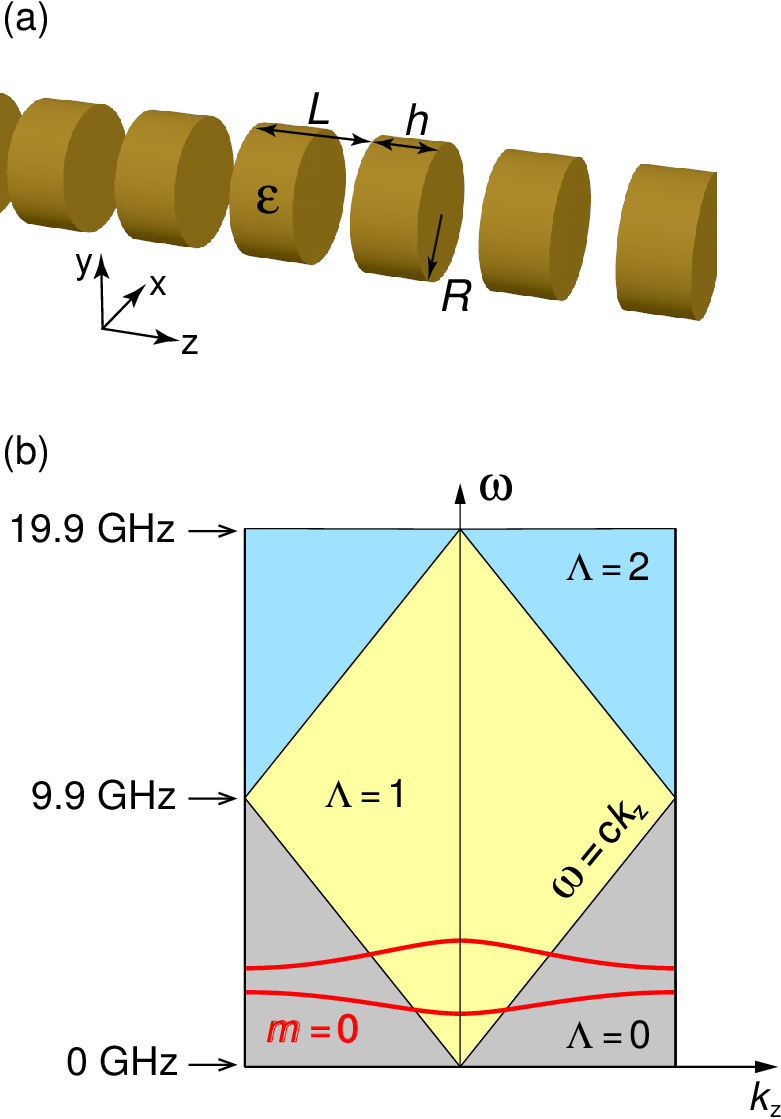}
\caption{(a) Chain of ceramic disks with $\varepsilon=40$, radius $R=10.2$~mm and thickness $h=10.1$ mm arranged in an array with period $L=15.1$~mm. (b) Diagram showing the number of open diffraction channels, $\Lambda$, depending on frequency $\omega$ and Bloch wave vector $k_z$.} \label{fig1}
\end{figure}
%---------------------------------------------------------------------------------------------------(1)
%\begin{equation}\label{Hr Hz}
%E_{\phi}=\psi_{_{TE}}, ~~ H_r=\frac{i}{k_0}\frac{\partial
%\psi_{_{TE}}}{\partial z}, ~~ H_z=-\frac{i}{k_0r}\frac{\partial
%r\psi_{_{TE}}}{\partial r},
%\end{equation}
%other components of EM field equal zero, where the function $\psi_{TE}$ obeys equation
%%---------------------------------------------------------------------------------------------------(2)
%\begin{equation}\label{TE}
%    \left[\frac{\partial^2}{\partial r^2}+\frac{1}{r}\frac{\partial }{\partial r}
%    -\frac{1}{r^2}+\frac{\partial^2}{\partial z^2}+U_{_{TM}}(z)\right]\psi_{_{TE}}(z,r)=0,
%\end{equation}
%%---------------------------------------------------------------------------------------------------(3)
%\begin{equation}\label{UTE}
%    U_{TE}(z)=\epsilon(z)\omega^2.
%\end{equation}

\begin{figure*}[t]
\includegraphics[width=2\columnwidth]{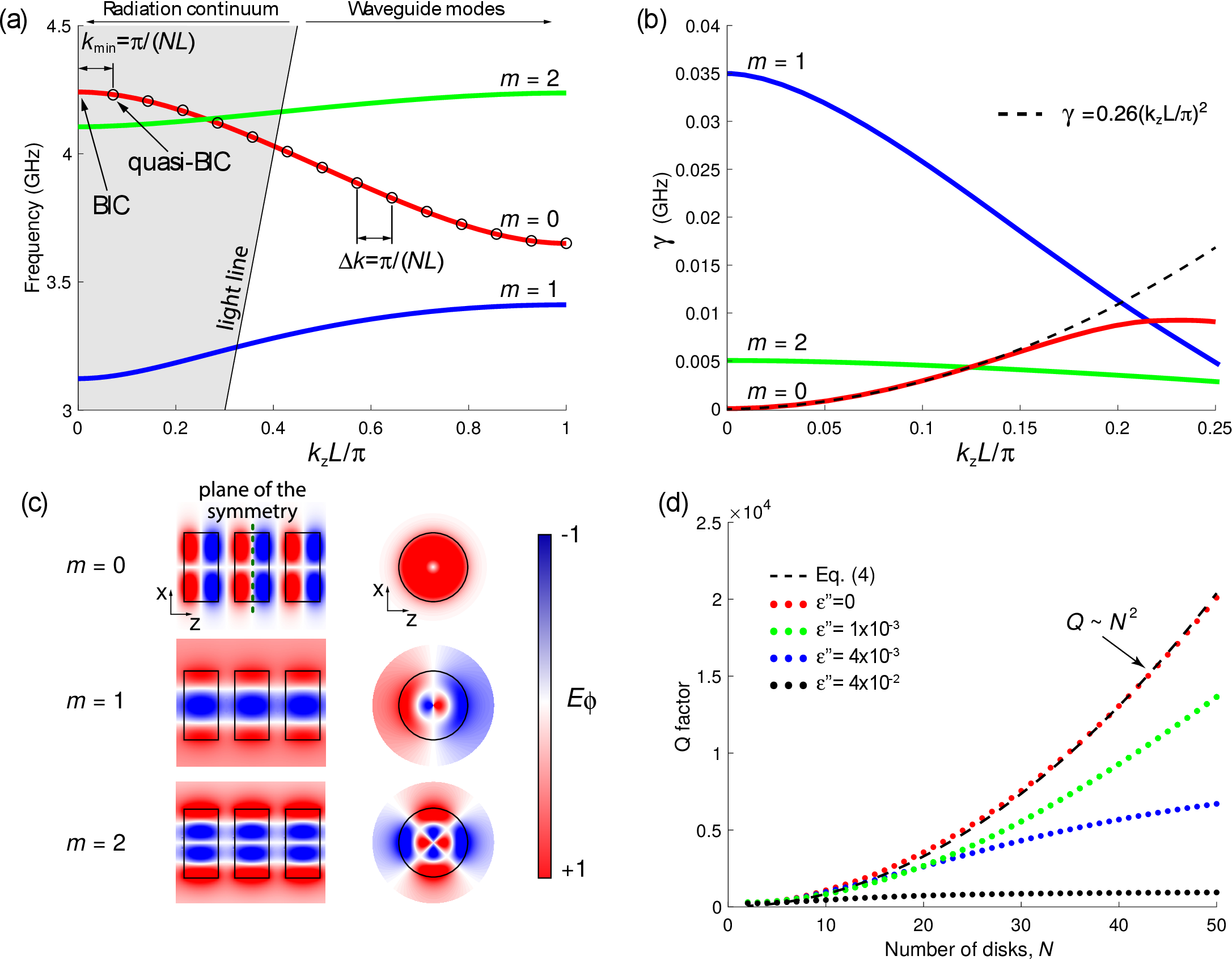}
\caption{(a) Dispersion of three eigenmodes with OAM $m=0,1,2$ in the infinite chain of the ceramic disks. The $\Gamma$-point for the mode with $m=0$ corresponds to BIC.  The parameters of the chain is mentioned in the caption of Fig.~\ref{fig1}. The black circles show the dispersion of eigenmode with $m=0$ in a finite chain consisting of $N$ periods. (b) Dispersion of the radiation losses for eigenmodes with OAM $m=0,1,2$ in the infinite chain of the ceramic disks.   (c) Side and front views of the distribution of the azimuthal electric
field for the eigenmodes with OAM $m=0,1,2$ at the $\Gamma$-point. The upper row shows the field distribution for BIC. (d) Dependence of Q factor on the number of periods in the chain for different level of material losses obtained numerically in Comsol Multiphysics. Dashed line is the quadratic approximation given by Eq.~\eqref{eq:approx_1}.} \label{simul1}
\end{figure*}

In the considered geometry, the variable $z$ and $r$ are not separable. Therefore, the eigenfunctions $\mathbf{U}_{m,n,k_z}(z,r)$ and dispersion of the modes could be found only numerically. Figure~\ref{simul1}(a) shown the dispersion of three eigenmodes with OAM $m=0,1,2$ calculated numerically in Comsol Multiphysics.
The parameters of the chain {are listed} in the caption of Fig.~\ref{fig1}.  Below the light line, the modes have no radiation losses. Above the light line the modes are leaky {due to emanation into the first open diffraction channel}. The dispersion of the radiation losses characterized by $\gamma=\text{Im}\left[\omega(k_z)\right]$  is shown in Fig.~\ref{simul1}(b). One can see $\gamma$ remains constant in the $\Gamma$-point for the modes with $m=1,2$.  However, the radiation losses  for the mode with $m=0$ tends to zero quadratically with $k_z$ in the vicinity of the $\Gamma$-point. Therefore, the mode with $m=0$ turns into the BIC in the  $\Gamma$-point. 
The statement that this state is symmetry protected immediately follows from the distribution of the electric field $E_\varphi$ [see Fig.~\ref{simul1}(c)]. It is antisymmetric with respect to the center plane of the disk, and therefore it vanishes after averaging over the unit cell.

The states with $m=0$ are qualitatively differs from the states with nonzero OAM.  Indeed, polarizations of the modes with $m=0$ is completely separable into TE and TM. For TE modes $\mathbf{E}=(0,E_\varphi,0)$ and   $\mathbf{H}=(H_r,0,H_z)$ and  for TM modes  $\mathbf{E}=(E_r,0,E_z)$ and $\mathbf{H}=(0,H_\varphi,0)$. The BIC with $m=0$ shown in Fig.~\ref{simul1}(a) can be classified as a TE mode. 
This classification remains  valid even for finite chains. Below, in Sec.~\ref{sec:transmission}, we will use this fact for selective excitation of the symmetry protected BIC with $m=0$ via near fields.

%One can see that only the mode with $m=0$ turns into the symmetry protected BIC at the $\Gamma$-point. Its radiation losses \Almas{tend} to zero quadratically and the losses of other modes remain finite at the $\Gamma$-point.

%The electric field profiles for all the considered modes at the $\Gamma$-point are shown in Fig.~\ref{simul1}(c). One can see that the field distribution of the symmetry protected BIC is antisymmetric with respect to the \Almas{center plane} of the disk, and therefore its averaging over the unit cell gives zero.

%Each band in the chain could be associated with a mode of a single dielectric disk. The mode resulting in appearance of BIC is TE_$0,$

 %explained in as  The emergence of the bands in the chain

%\comment{each mode can be associated with resonance of solitary disk}

%--------------------------------------------------------------------------------------------Fig.2

\subsection{Finite chain}

In practice, we always deal with finite arrays of scatterers. Therefore, the diffraction continuum is not quantized completely, and the diffraction channels are smeared, providing the radiation in the range of angles around diffraction directions if the infinite chain. Strictly speaking, this results in destroying of a true BIC and turns it into a quasi-BIC -- a resonant state with high but finite value of Q factor.

The dependencies of Q factor on the number of scatterer, $N$, for arrays of different dielectric particles have been studied theoretically for waveguide modes, leaky modes \cite{Blaustein,Polishchuk}, and quasi-BIC~\cite{AdvEM,Peng,NiOE2017,BulgMaksOE,Taghizadeh}. In particular, it was predicted that Q factor of the symmetry protected BIC is proportional to $N^2$ while the Q factor of accidental BICs is proportional to $N^3$.  Our simulations confirm that in the absence of material losses, the Q factor of the quasi-BIC at the $\Gamma$-point grows as $N^2$ [Fig.~\ref{simul1}(d)]. The experimental study of this dependence is provided is Sec.~\ref{sec:transmission}. 

The dependance of Q factor on $N$ can be obtain from the known dependence of $\gamma(k_z)$ in the infinite chain [see dashed line in Fig.~\ref{simul1}(b)]. Indeed, applying the Born-Karmen boundary conditions to a finite chain it is easily to get that all resonances in the chain placed equidistantly in the Brillouin zone with the distance $\Delta k_z=\pi/(N L)$ [Fig.~\ref{simul1}(a)]. The resonance with minimal Bloch wave vector $k_z^{\text{min}}=\Delta k_z$ corresponds to the quasi-BIC.  Substitution of $\Delta k_z$ into $\gamma(k_z)$ gives a simple estimation for the dependence of the Q factor of the quasi-BIC on the number of periods: 
\begin{equation}
Q(k_z^{\text{min}})=\frac{\omega_{\text{BIC}}}{2\gamma(\Delta k_z)}=\frac{4.24}{2\cdot 0.26}N^2\approx 8.15 N^2.
\label{eq:approx_1}
\end{equation}
{Surprisingly, this estimation well agrees} with the results of numerical simulation [see Fig.~\ref{simul1}(d)]. This result means that radiation due to scattering on the ends of the chain is negligible.

Besides the radiation losses due to a finite size of the chain, there {are other sources of losses} (absorption in material, roughness of the scatterers, structure disorder, leakage into high index substrate etc.) which contribute to the total losses, even in the case of the infinite chain~\cite{AdvEM,Peng,NiOE2017,Sadrieva}.  The practically important question is how many disks it is necessary to take to ensure that the radiation losses due to the finite size of the sample will be negligible with respect to other losses.
Figure~\ref{simul1}(d) shows the dependence of the Q factor of the quasi-BIC on the number of disks $N$ for different level of material losses introduced through imaginary part of the permittivity $\varepsilon''$. {For larger} number of the disks the total Q factor saturates. It means that material losses give the main contribution into the total losses.  {The maximal achievable Q factor could be estimated as
$Q_{\text{max}}=\varepsilon'/\varepsilon''$.}

\begin{figure*}[htbp]
\includegraphics[width=1\linewidth]{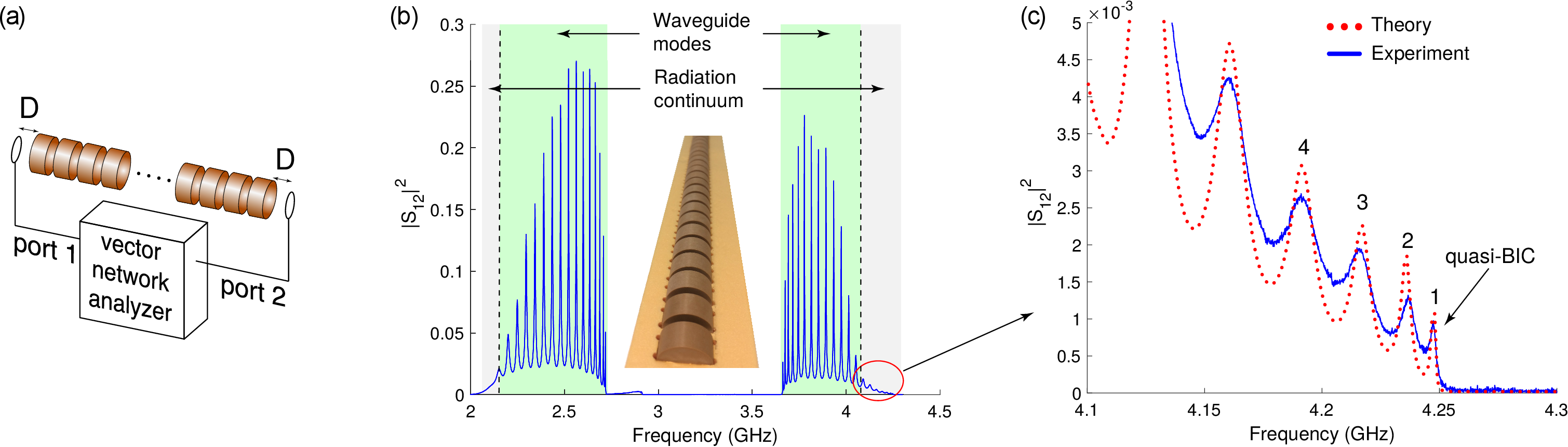}
\caption{(a) Artistic view of the experimental setup for measurement of the transmission spectra of the chain of the ceramics disks.  (b)~Transmission spectra of the chain consisting of 20 ceramics disks placed between two coaxially positioned loop antennas. The parameters of the chain are shown in caption of Fig.~\ref{fig1}. The green and grey areas correspond to the waveguide and leaky modes, respectively. The inset shows the photo of the sample.  (c) Zoomed-in view of the transmission spectra shown in panel (b). The dotted line shows the results of numerical simulations carried out in Comsol Multiphysics. The last peak in the series corresponds to quasi-BIC.}
\label{simul5}
\end{figure*}

\section{Sample}

The sample represents a finite chain of the disks  fabricated from BaO-TiO$_2$ microwave ceramic placed equidistantly with the period $L$=15.1 mm (see Fig.~\ref{fig1}). The permittivity of the ceramics and tangent of losses extracted from the auxiliary experiment on measurement of the scattering cross-section on a single ceramics disks in microwave frequency band (2-5 GHz) are equal to $\varepsilon=40.0$ and $\tan \delta = 2.5\times 10^{-4}$,
respectively ({see Sec.~S1 in Supplemental Material}). The radius and the height of the disks are $D$=10.2~mm and $h$=10.1~mm,  respectively. To fix the array a special holder with groves was fabricated from a Styrofoam material with a permittivity close to 1.1 at the microwave frequencies. Simple calculations show that the holder results in blueshift of the BIC with $m=0$ at $\Gamma$-point less than 1 MHz and it does not affect the radiation losses at all since the BIC is symmetry protected. 

The array of the disks has two main advantages over the array of spheres. The first one is that disk resonators are easer to fabricate using a conventional sintering procedures of microwave ceramic powder after pressing it in a steel die. The second one is that disks have two scale parameters, height $h$ and radius $R$. Their independent variation together with the period of the chain $L$ allows to provide precise mode engineering getting a number of BICs with different orbital angular momenta and Bloch
vectors \cite{PRA96}.

%--------------------------------------------------------------------------------------------Fig.5
\begin{figure}[t]
\includegraphics[width=1\columnwidth]{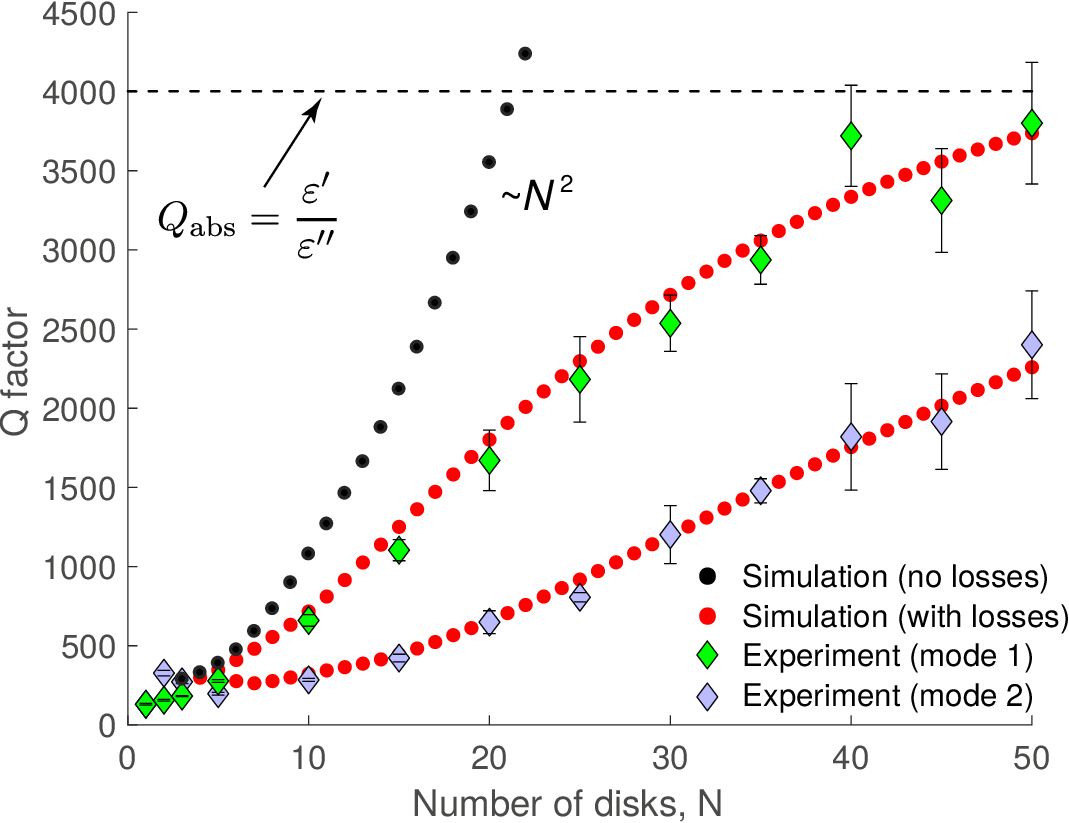}
\caption{Experimental dependence of Q factors of the symmetry protected quisi-BIC (mode 1) and the neighboring resonance (mode 2) on the number of disks $N$ in the chain of ceramics disks. 
Error bars indicate the standard deviation in the Q factor extracted from the transmission spectra measured five times. The parameters of the chain is mentioned in the caption of Fig.~\ref{fig1}. Dotted line shows the results of numerical simulation in Comsol Multiphysics.} \label{simul4}
\end{figure}

It follows from Fig.~\ref{simul1} that the frequency bands for $m=0$ and $m=2$ overlaps. This hinders the observation of the symmetry
protected BIC through measurement of the scattering cross-section since the incident plane wave is contributed by cylindrical waves with all OAM.
Thus, our preliminary experiments on measurement of scattering cross-section of the chain provided insufficient results because we cannot clearly distinguish the
modes with $m=2$ and quasi-BIC with $m=0$ ({see Sec.~S2 in Supplemental Material}).

%--------------------------------------------------------------------------------------------Fig.7
\begin{figure*}[htbp]
\includegraphics[width=1\linewidth]{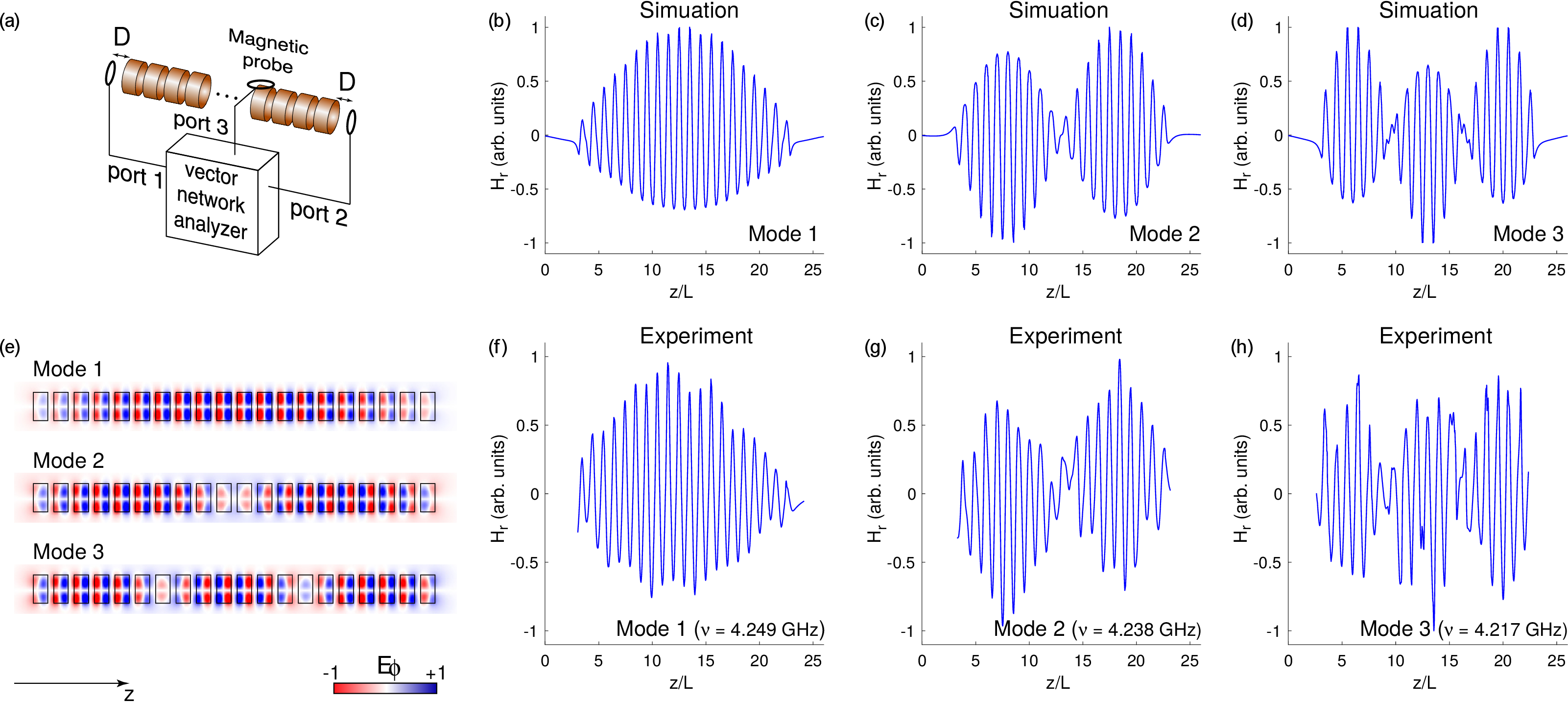}
\caption{(a) Artistic view setup for measurement of the distribution of the radial magnetic field component. (b)-(d) Numerically obtained distribution of the radial magnetic field component for the chain consisting of 20 ceramics disk. (e) Numerically obtained distribution of the azimuthal electric field component for the chain consisting of 20 ceramics disk. (f)-(h) Measured distribution of the radial magnetic field component for the chain consisting of 20 ceramics disk.} \label{simul6}
\end{figure*}

\subsection{Transmission spectra \label{sec:transmission}}

As we mentioned above (see Sec.~\ref{sec:inf_chain}), the polarization of the mode with $m=0$ is separated into TE and TM. Therefore, they can be selectively excited via near field by axially symmetric antennas placed coaxially with the chain. The TM modes can be excited by an electric dipole antenna, and the TE modes  can be excited by a magnetic dipole antenna. Since the analyzed symmetry protected BIC is a TE-mode, in the experiment we use two identical shielded-loop antennas~\cite{Whiteside} placed coaxially with the chain and connected to ports of a vector network analyzer (VNA) [see Fig.~\ref{simul5}(a)].  The antennas with the outer diameter of 10 mm have been fabricated from 086 Semi-rigid Coax Cable.
They are placed at the distance $D=5$~mm away from the faces of the first and last disks. Such a distance provides a weak coupling regime
between the antennas that makes the analysis of the Q factor of the quasi-BIC eligible ({see Sec.~S3 in Supplemental Material}).
%Thus, the antennas do not decrease the Q factor of the quasi-BIC substantially.

%To measure the transmission spectra the array of the disks has been loaded from the edges by two identical shielded-loop antennas connected to the first and second ports of the VNA. The shielded-loop antennas~\cite{Whiteside} with the outer diameter of 10 mm have been fabricated from 086 Semi-rigid Coax Cable. To measure the transmission coefficient the antennas have been placed 5 mm away from the edges of the disks array symmetrically to the array axis.

The measured transmission spectra of the array of 20 disks placed between two loop antennas is shown in Fig. \ref{simul5}(b). Two transmission bands
consisting of 20 resonance peaks each are clearly seen. These bands corresponds to the modes with $m=0$ shown in Fig.~\ref{fig1}(b).
A weak ripples at frequencies 2.7-2.9~GHz corresponds to the modes with $m=1$, which are excited due to non-perfect axial symmetry of the sample.
The resonances laying in the green area in Fig. \ref{simul5}(b) corresponds to the waveguide modes of the infinite chain and the resonances
in the grey corresponds to the leaky modes at the infinite chain. The intensity of the leaky resonances in the transmission spectrum is very weak because of coupling with the continuum of radiation modes in surrounding space.
%In the sector with zero OAM $m=0$ we have two branches one of which lies below the light line. The Q factor of each mode from this branch is determined mainly by contribution of material losses of disks. While part of the eigenmodes emerges into the radiation continuum as shown in Fig. \ref{simul1} and dash line in Fig. \ref{simul2} that crucially damages the Q-factor of these modes as seen from the transmission spectra in Fig. \ref{simul5}. where the each
Panel~(b) in Fig.~\ref{simul5} shows the zoomed transmission spectra in the region of leaky modes. The blue solid line is the experimental
data and the red dotted line is the results of numerical simulation carried out in Comsol Multiphysics. Figure \ref{simul5} shows clearly that
the width of the last peak in the series is the most narrow. This peak corresponds to quasi-BIC, which transforms into a true BIC
as $N\rightarrow \infty$.

Each transmission spectra with a fixed number of the disks is measured five times. After each measurement, the disks are extracted from the holder and shuffled. The experimental dependence of the Q factor for two last resonances in the series [see mode 1 and mode 2 in Fig.~\ref{simul5}(c)] on the number of the disks is shown in Fig.~\ref{simul4} by rhombus markers. The error bars show the standard deviation in the Q factors.  The dotted lines correspond to the simulation.   One can see that for small $N$, when the radiative losses are dominant, the Q factor increases quadratically. However, the deviation from the quadratic {behavior} becomes essential as $N\sim20$. Further increase of the number of the disks results in saturation of the total Q factor to the level $Q_{\rm abs}=\varepsilon'/\varepsilon''$. {It is clear that} the lower the material losses the bigger number of scatterers {necessary to take in order} {is needed} to suppress the radiative losses of quasi-BIC with respect to the material absorption.
The analyzed chain of the ceramics disks with  $\tan \delta = 2.5\times 10^{-4}$ behaves as an infinite one when the number of the disks is more than 50.

%For the analyzed ceramics with $\tan \delta = 2.4\times 10^{-4}$, the chain consisting of about 50 disks could be considered as an infinite chain.

%in spite of each transmission peak has a finite width because of material losses the last peak is the most narrow. Exactly this peak is responsible for the symmetry protected BIC in the limit $N\rightarrow \infty$ if there were no material losses or other factors discussed above. The corresponding Q factor inverse to these resonant widths behave as shown in Fig. \ref{simul4}.

%\comment{Last peak is quasi-BIC}
%
%\comment{The extracted dependence of Q-factor + theory}
%
%\comment{Quadratic growth at small $N$, then it saturates}
%
%\comment{Discussion on inaccuracy}
%
%\comment{How many disks do we need. Estimation ....}

\subsection{Measurements of mode profiles}

In the previous section we analyzed the spectral characteristics of quasi-BIC. Here, we focus on the experimental study of its field profile.
%\comment{To be sure that we resolved all resonances in the transmission spectra and indeed analyze the transmission of ...... extract }
A true symmetry protected BIC with $m=0$ being TE-polarized mode has only three nonzero components of electromagnetic field, $E_\varphi$, $H_r$, and $H_z$. It is notable that for the considered mode, the component $H_r$ is not equal to zero only for the closed diffraction channel. Therefore, it characterized only the structure of near field.  To measure the radial component of magnetic field $H_r$  we  use the shielded-loop antenna as a probe [see Fig.~\ref{simul6}(a)]. The antenna of the same geometry as the previous two has been connected to the third port of the VNA and it fixed to an arm of a high-precision scanner equipped by automated step motor. The magnetic field was probed at the distance of 1 mm above the disks along the array with the step of 0.5 mm. The frequency range to scan the magnetic field has been determined by the transmission coefficient measurement.

The measured magnetic field profiles for quasi-BIC and two neighbour resonances [see peaks 2 and 3 in Fig.~\ref{simul5}(a)]  for the chain
of 20 disks are shown in Figs.~\ref{simul6}(f,g,h). The results of the numerical simulations carried out in Comsol Multiphysics are shown
in~Figs.~\ref{simul6}(b,c,d). It is clear seen that for the mode at $f=4.249$~GHz the envelope of the filed profile is a half of the sine
period. Therefore, the observed mode is a symmetry protected quasi-BIC with the minimal possible Bloch wave vector equal to $k_z=\pi/(LN)$.

\begin{figure}[t]
\includegraphics[width=1\columnwidth]{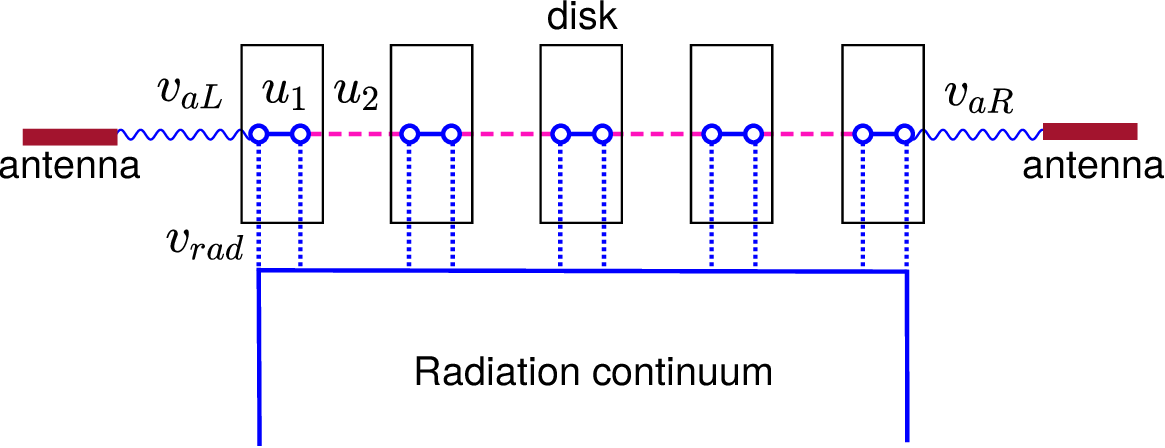}
\caption{The tight-binding simulation of the N disks coupled with the left and right antennas via the couplings $v_{aL}$ and
$v_{aR}$. Whole chain is coupled with wide waveguide simulating the radiation continuum via the coupling $v_{rad}$.} \label{tb}
\end{figure}
%--------------------------------------------------------------------------------------------Fig.9

\begin{figure}
\includegraphics[width=1.0\linewidth]{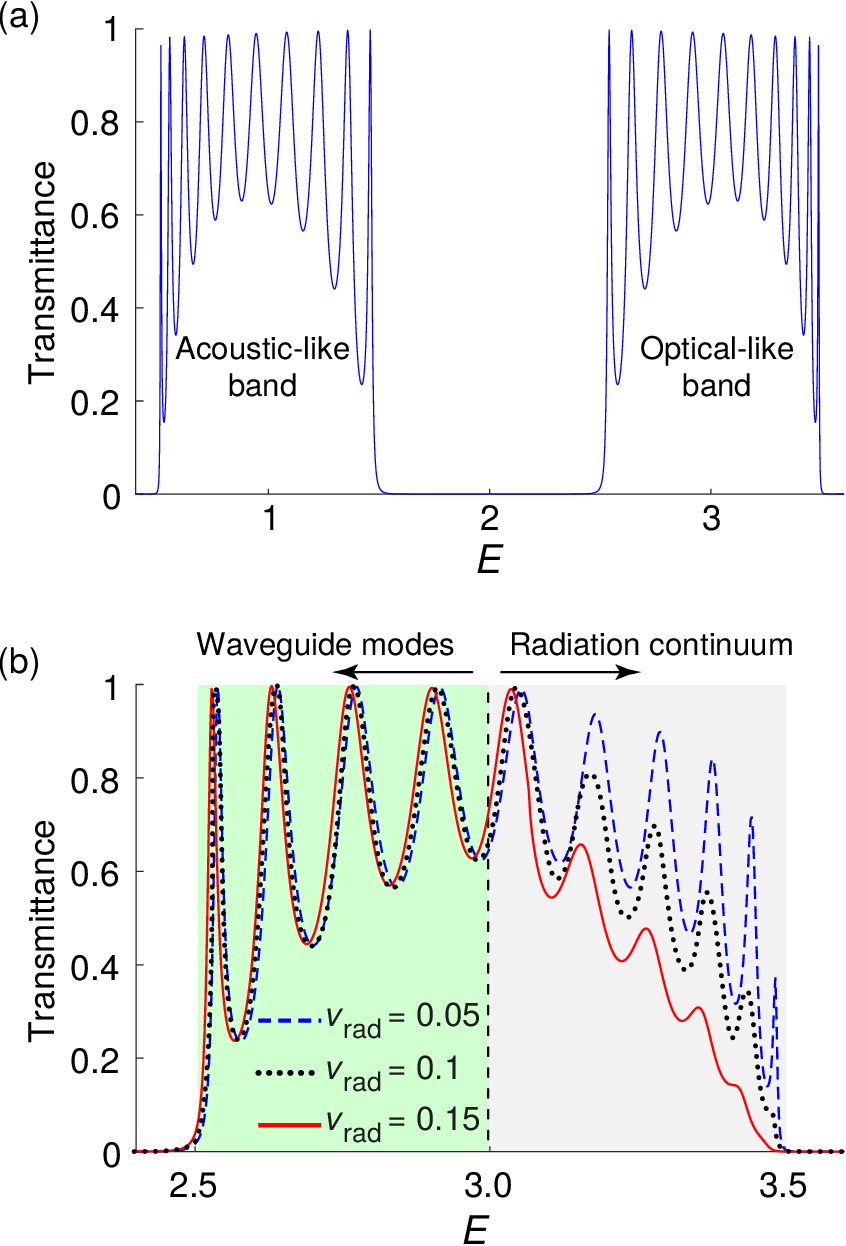}
\caption{The transmission spectra of the tight-binding model for (a) $v_{rad}=0$ and (b) for $v_{rad}\neq 0$, $u_1=0.5, u_2=1, v_{aL}=0.5, v_{aR}=0.6$.} \label{vrad}
\end{figure}

\section{Tight-binding model}
It is clearly seen from Fig.~\ref{simul5}(b) that the transmission drops dramatically for resonances above the light line. It happens due to their coupling with radiation continuum. This behaviour could be describe in the framework of tight-binding approach accounting for interaction between the neighbour disks and their coupling to the radiation continuum. The use of tight-binding approach is natural because the ceramics disks composing the chain have a permittivity $\varepsilon=40$ that makes the field well localized inside the disks [see Figs.~\ref{simul1}(c) and~\ref{simul6}(e)].   

The tight-binding Hamiltonian of the chain consisting of $N$ disk has $2N$ nodes coupled by alternated way as shown in Fig.~\ref{tb} ~\cite{SR}:
%The transmission spectra shown in Fig. \ref{simul5} is similar to the spectra through the chain consisted of $2N$ sites with the Hamiltonian
\begin{equation}\label{tb}
    \widehat{H}_B=\sum_{j=1}^{2N}|\psi_j|^2-\sum_{j=1}^{2N-1}u_j\psi_j\psi_{j+1}^{*}+h.c.
\end{equation}
Therefore, each disk is described by a pair of nodes with coupling coefficient $u_1$. Two nodes per disk is needed to realize an antisymmetric mode of the disk resulting in appearance of BIC in the chain. The coupling between the disks is described by coefficient $u_2$. The effective Hamiltonian accounting for radiation losses and coupling with antennas can be written as follows:
\begin{equation}\label{tbeff}
    \widehat{H}_{eff}=\widehat{H}_B-\sum_{j=1}^{2N}[v_{aL}e^{ik}\delta_{j,1}+v_{aR}e^{ik}\delta_{j,2N}+v_{rad}e^{ik_p}].
\end{equation}
Here, the coefficients $v_{aL}$ and $v_{aR}$ is responsible for coupling with transmitting and receiving antennas, respectively, with the propagation band
\begin{equation}\label{k}
    E=4\sin^2k.
\end{equation}
The coefficient $v_{rad}$ is responsible for coupling with radiation continuum, which could be considered as a wide waveguide coupled to the array along its entire length. In order to only a part of eigenvalues were embedded into the
radiation continuum we shift the cutoff energy of the radiation continuum separating spectra of leaky and waveguide modes to the level $E=3$:
\begin{equation}\label{prop}
    E=3+8\sin^2 k_p.
\end{equation}

The reference transmission spectrum for $N=10$ calculated neglecting the coupling with radiation continuum ($v_{rad}=0$)  is shown in Fig.~\ref{vrad}(a). The used parameters are listed in the caption. One can see that the spectrum consists of the acoustic-like and optical-like bands that agrees with the experimental data [see Fig.~\ref{simul5}(b)]. The transmission at the resonances is close to unit. Introduction of the coupling ($v_{rad}\neq0$) with radiation continuum does affect only the modes with energies above the cutoff [Fig.~\ref{vrad}(b)]. The coupling to the continuum results in a leakage of pumped wave from left antenna into the radiation continuum and respectively a decrease of the transmission peaks at resonances energies above embedded into radiation continuum. The following increase of $v_{rad}$ results in blurring of the transmission peaks. The considered toy model  well describes the experimental behaviour of the transmission shown in Figs.~\ref{simul5}(b) and~\ref{simul5}(c).

\section{conclusion}

In this paper we have observed a symmetry protected bound state in the continuum (BIC) in the linear array of  periodically arranged ceramic disks for the first time. In the experiment, we selectively excited the modes with zero orbital angular momentum by coaxially placed loop antennas and measured the transmission spectra of the array.  We analyzed the dependence of the Q factor of the BIC on the number of disks in the chain and estimated the critical number of the disk making the radiation losses become negligible with respect to the material absorption. For the considered ceramics with tangent of losses $\tan\delta=2.5\times10^{-4}$ this critical number of the disks is about 50. We confirmed the observation of BIC by the measurements of the magnetic field profiles.  All measurements are in a good agreement with the results of numerical simulation and analytical model based on tight-binding approximation. The obtained results provide useful guidelines for practical implementations of structures with bound states in the continuum that opens up new horizons for the development of optical and radiofrequency metadevices.  

%We presented the first report of experimental observation of the symmetry protected BIC in the linear array of  periodically arranged ceramic disks. Due to the Styrofoam material of the cuvette with permittivity close to an air the system turns out to be close to the system considered in \cite{PRA96} with however a few important aspects which makes difference between theory and experiment. The first aspect is a finite number of disks which altogether with material losses and structural fluctuations brings on agenda a concept of quasi BICs as resonant states with very high Q factor. The measurement of this Q factor was a subject of the paper. The second aspect is related to overlapping of eigenmode subbands of the array which belong different OAM with $m=0, 1, 2$. That aspect complicates direct observation of quasi BICs in scattering of plane waves as was considered in theory \cite{PRA2018}. Nevertheless we are going to resume such kind of experiments for another more favorite choice of disks lengths. However observation of the non symmetry protected BICs with OAM is our high priority.

\acknowledgements 
The numerical simulations and experimental part of this work were supported by the Russian Science Foundation (17-12-01581), the analytical calculations were supported by
RFBR (16-02-00314). The authors are thankfull to Andrey Sayanskiy for an assistance in holder fabrication. Mikhail Balezin and Mikhail Belyakov contributed equally to this work.

\bibliography{references-1}

\end{document}